\documentclass[twoside]{llncs}
\usepackage{amsmath}
\usepackage{amssymb}
\usepackage{graphics}
\usepackage[dvips]{epsfig}
\usepackage{times}
\usepackage{algorithm}
\usepackage{algorithmic}
\usepackage{fancyhdr}           
\usepackage{ifthen}
\newcommand\settitle[2][]{%
 \title{#2}
 \ifthenelse{\equal{#1}{}}%
  {\fancyhead[RO]{\nouppercase #2 \qquad \thepage}}%
  {\fancyhead[RO]{\nouppercase #1 \qquad \thepage}}%
}
\newcommand\setauthors[2]{%
 \author{#2}
  {\fancyhead[LE]{\thepage \qquad \nouppercase #1}}%
}
\fancypagestyle{plain}{%
\fancyhf{}

}

\def\keywordsname{Keywords.}
\newenvironment{keywords}{%
      \list{}{\advance\topsep by-0.50cm\relax\small
     \leftmargin=1cm
      \labelwidth=1cm
     \listparindent=1cm
     \itemindent\listparindent
      \rightmargin\leftmargin}\item[\hskip\labelsep
                                    \bfseries\keywordsname]}

   {\endlist}

   \newtheorem{Sublem}{Sublemma}[section]

   \def\C{{\mathbb C}}
   \def\R{{\mathbb R}}
   \def\N{{\mathbb N}}

   \def\Pf{{\it Proof.$\;\;$}}

   \def\qed{\hfill$\diamond$}

   \def\cB{{\mathcal B}}
   \def\cM{{\mathcal M}}
   \def\cQ{{\mathcal Q}}

   \def\cP{{\mathcal P}}

   \def\ov{\overline}
   \def\ovH{\ov{H}}
   
   \def\unH{\underline{H}}

\begin{document}

\settitle[]
         {On analytic properties of entropy rate}

\setauthors{A. Sch\"onhuth}
           {Alexander Sch\"onhuth}
\institute{Pacific Institute for the Mathematical Sciences\\
           School of Computing Science\\
           Simon Fraser University\\
           8888 University Drive\\
           Burnaby, BC, V5A 1S6, Canada\\
\email{schoenhuth@cs.sfu.ca}}

\date{}
\maketitle

\thispagestyle{plain}
\begin{abstract}
Entropy rate is a real valued functional on the space of discrete
random sources for which it exists. However, it lacks existence proofs
and/or closed formulas even for classes of random sources which have
intuitive parameterizations. A good way to overcome this problem is to
examine its analytic properties relative to some reasonable
topology. A canonical choice of a topology is that of the norm of
total variation as it immediately arises with the idea of a discrete
random source as a probability measure on sequence space. It is shown
that both upper and lower entropy rate, hence entropy rate itself if
it exists, are Lipschitzian relative to this topology, which, by well
known facts, is close to differentiability. An application of this
theorem leads to a simple and elementary proof of the existence of
entropy rate of random sources with finite evolution dimension. This
class of sources encompasses arbitrary hidden Markov sources and
quantum random walks.
\end{abstract}


\section{Introduction}
Entropy rate is a key quantity in information theory
as it is equal to the average amount of information per symbol of
discrete-time, discrete-valued stochastic processes (usually referred
to as {\em discrete random sources} in the following).  Therefore, it
is natural to ask how entropy rate behaves if knowledge of discrete
random sources is subject to uncertainties which, for example, may be
inherent to inference processes and/or originate from noisy
channels. However, closed formulas for entropy rate exist only for
rare examples of classes of discrete random sources.  For instance,
already hidden Markov sources (HMSs) seem to defy a convenient formula
although there is one for the special case of Markov sources.
Therefore, in this case, recent efforts focused on the direct
investigation of analytic properties of entropy rate like smoothness
or even analyticity \cite{Marcus06,Marcus07}, \cite{Zuk05,Zuk06},
\cite{Ordentlich06}, \cite{Jacquet04}.

\medskip 
The purpose of this paper is to contribute to the issue of analytic properties
of entropy rate in a more general fashion. Namely, we study the behavior of
entropy rate relative to the topology induced by the norm of total
variation. This topology is one of the natural choices and it is ubiquitous in
both theoretical and practical work. We show that entropy rate is
Lipschitzian on the whole space of discrete random sources which is, due to an
elementary theorem of Rademacher, close to differentiability.\par We will use
this result to give an elementary proof of the existence of entropy rate for
sources with finite evolution dimension \cite{Faigle07} which contain the
classes of arbitrary HMSs \cite{Merhav02} and quantum random
walks (QRWs)\cite{Aharonov01}, \cite{Faigle06}.

\medskip
The paper is organized as follows. We will identify discrete random sources
with probability measures acting on the measurable space of symbol sequences
equipped with the $\sigma$-algebra generated by the cylinder sets of
sequences.  Therefore, in section \ref{sec.prelim}, we will briefly compile
the theory's standard arguments. In section \ref{sec.analytic} we prove that
entropy rate is Lipschitz continuous relative to the topology induced by the
norm of total variation which is the main contribution of this paper. In
section \ref{sec.finitedim} we demonstrate how to exploit this result for an
elementary proof of existence of random sources with finite evolution
dimension which include HMSs and QRWs as special cases. In section
\ref{sec.conclusion} we will describe the proof's intuition thereby commenting
on open problems such as other choices of topology and/or stricter choices of
analytic properties.

\section{Random sources and entropy rate}
\label{sec.prelim}

As usual, $\Sigma^*=\cup_{t\ge 0}\Sigma^t$ is the set of all words
(strings of finite length) over the finite alphabet $\Sigma$ together
with the concatenation operation
\begin{equation}
v\in\Sigma^t, w\in\Sigma^s
\quad\Rightarrow\quad vw\in\Sigma^{t+s}.
\end{equation}
Throughout this paper $\Omega=\Sigma^{\N}=\bigotimes_{t=0}^{\infty}\Sigma$ is
the set of sequences over $\Sigma$ and $\cB$ is the $\sigma$-algebra generated
by the cylinder sets. Cylinder sets $B$ are identified with sets of words
$A_B\subset\Sigma^t$ such that $B$ is the set of sequences which start with the
words in $A_B$. In general, the cardinality of a set $A$ is denoted by
$|A|$.\\

We view stochastic processes $(X_t)_{t\in\N}$ with values in $\Sigma$
as probability measures $P_X$ on the measurable space $(\Omega,\cB)$
and vice versa via the relationship
($v=v_0...v_{t-1}\in\Sigma^t$ corresponds to the cylinder set of
sequences having $v$ as prefix)
\begin{equation}
\label{eq.stringfunction}
P_X(v) = P(\{X_0 = v_0,X_1 = v_1,...,X_{t-1} = v_{t-1}\}),
\end{equation}
where the term on the right hand side is the probability that the
random source emits the symbols $v_0,...,v_{t-1}$ at periods
$0,...,t-1$.  Note that
a stochastic process $(X_t)$ is uniquely determined by the values
$P_X(v)$ for all $v\in\Sigma^*$ as the cylinder sets
corresponding to words $v$ generate $\cB$\\

Although being a canonical choice of norm (see appendix
\ref{app.totvarnorm} for a short review of the related theory and
corresponding definitions), computation of the norm of total variation
would not be easy for the measurable space under consideration by
means of its original definition alone. The following lemma shows a
concrete way to get a grip of the corresponding topology. Exact
definition and basic properties of the norm of total variation have
been deferred to appendix~\ref{app.totvarnorm}.\\

\begin{lemma}
\label{l.metric}
The topology induced by the norm of total variation is that
of the metric
\begin{equation}\label{eq.totvar}
d_{TV}(P_X,P_Y) = \sup_{t\in\N}\sum_{v\in\Sigma^t}|P_X(v)-P_Y(v)|
                    = \lim_{t\to\infty}\sum_{v\in\Sigma^t}|P_X(v)-P_Y(v)|.
\end{equation}
where $P_X,P_Y$ are probability measures associated to 
discrete random sources $(X_t),(Y_t)$. 
\end{lemma}

\Pf
See sec.~\ref{ssec.metricproof} of the appendix for the predominantly
measure theoretical arguments.
\qed\\

\subsection{Entropy Rate}
\label{ssec.entropyrates}

In the following, we will refer to the quantities 
\begin{eqnarray}
\label{eq.entropydef}
\ovH(X) & := & \ovH(P_X) := \limsup_{t\to\infty} H^t(P_X)\\
\text{resp.}\quad\unH(X) & := & \unH(P_X) := \liminf_{t\to\infty} H^t(P_X)
\end{eqnarray}
as {\em upper entropy rate} resp. {\em lower entropy rate} of a random
source $(X_t)$ with associated probability measure $P_X$, where, using
the language introduced above,
\begin{equation}
H^t(P_X) := -\frac1t\sum_{v\in\Sigma^t}P_X(v)\log\, P_X(v)
\end{equation}
is the entropy of the distribution over the words of length $t$
induced by the random source, divided by $t$. {\em Entropy rate}
of a random source $(X_t)$ with associated probability measure $P_X$
is denoted by
\begin{equation}
H(X) := H(P_X) := \lim_{t\to\infty}H^t(P_X).
\end{equation}
The existence of the limit of the $H^t(P_X)$ is also referred to as
the \emph{existence of entropy rate}
where, obviously, a necessary and sufficient condition for entropy rate to exist
is
\begin{equation}
\label{eq.entrateexistence}
\ovH(X) = \unH(X)\quad (= H(X)).
\end{equation}
Throughout this paper,
$\Delta^{n-1}=\{x=(x_1,...,x_n)\in\R^n\,|\,x_i\ge 0,\sum_ix_i = 1\}$
is the usual regular $n-1$-dimensional simplex in $\R^n$ and, for
technical convencience, $\log$ is the natural logarithm. Note that,
as it is more common to use the logarithm to the base $2$, switching
bases does not affect any analytic property of entropy rate.

\section{Analytic properties of entropy rate}
\label{sec.analytic}

Our main result is the following theorem, which states that entropy
rate is Lipschitz continuous with respect to the topology induced by
the norm of total variation. In the following let $\cP$ be the set of
the probability measures associated with discrete random sources,
viewed as a normed space. Elements of $\cP$ will be denoted by $P$ or
$Q$. We further denote the normed subspace of discrete random sources
for which entropy rate exists by $\cP_H$.

\medskip

\begin{theorem}[Lipschitz continuity of entropy rate]
\label{t.lipschitzentropy}
The real-valued functionals $\ovH$ and $\unH$ on $\cP$
are Lipschitzian with $Lip(\ovH) = Lip(\unH) = \log\,|\Sigma|$, that is,
for $P,Q\in\cP$,
\begin{eqnarray}
|\ovH(P) - \ovH(Q)| & \le & (\log\,|\Sigma|)\,d_{TV}(P,Q)\\
|\unH(P) - \unH(Q)| & \le & (\log\,|\Sigma|)\,d_{TV}(P,Q).
\end{eqnarray}

\end{theorem}

\medskip
Clearly, because of (\ref{eq.entrateexistence}),
a corollary of the theorem is that the same holds true for entropy
rate itself.
\medskip

\begin{corollary}
Entropy rate is Lipschitzian with $Lip(H) = \log\,|\Sigma|$, that is,
\begin{equation}
|H(P) - H(Q)| \le (\log\,|\Sigma|)\,d_{TV}(P,Q)
\end{equation}
where, here, $P,Q\in\cP_H$.
\end{corollary}

\medskip

We present two lemmata, which incorporate the essential ideas of the
proof of the theorem. We write
\begin{equation}
d_{TV,t}(P,Q) := \sum_{v\in\Sigma^t}|P(v)-Q(v)|.
\end{equation}
Lemma~\ref{l.metric} says that $\lim_{t\to\infty}d_{TV,t}(P,Q)=d_{TV}(P,Q)$.
Note that $d_{TV,t}$ is not a metric on $\cP$.

\medskip

\begin{lemma}\label{l.hnestimation}
Let $P,Q\in\cP$ such that $d_{TV}(P,Q)\le\frac1e$. Then
it holds that
\begin{equation*}
|H^t(P) - H^t(Q)|\\ 
\le (\log\,|\Sigma| + \frac1t\log\,\frac{1}{d_{TV,t}(P,Q)})\cdot d_{TV,t}(P,Q),
\end{equation*}
where $0\cdot\log\,\infty := 0$ in case of $d_{TV,t}(P,Q) = 0$.
\end{lemma}

\medskip
\noindent For the proof of this lemma we will need a technical sublemma.

\medskip
\begin{Sublem}\label{sl.xlnx}
Let $h(x):=x\log(1/x)$ for $x\in ]0,1]$ and $h(0)=0$.
Then, for $x,y\in [0,1]$,
\begin{equation}
|x-y| \le\frac{1}{e} \quad\Longrightarrow\quad |h(x)-h(y)| \le h(|x-y|).
\end{equation}
\end{Sublem}

\Pf 
Note first that $h'(x) = \log\frac{1}{x} - 1$ 
and $h''(x) = -\frac{1}{x}$. Hence $h$ is concave, has
a global maximum at $\frac{1}{e}$ and $h(\frac{1}{e})=\frac{1}{e}$.
Therefore 
$x\le h(x) \; \Leftrightarrow \; x\le\frac{1}{e}$ $(*)$.
Because of
\begin{equation}
\begin{split}
|h(x)-h(y)| &= |\;|h(x) - h(\frac{1}{e})| - |h(\frac{1}{e}) - h(y)|\;|\\
&\le \max\{ |h(x) - h(\frac{1}{e})|, |h(\frac{1}{e}) - h(y)|\} \\
\end{split}
\end{equation}
and the fact that $h$ is monotonically increasing on $[0,\frac{1}{e}]$ we can,
without loss of generality, assume that either $x,y\ge \frac{1}{e}$ or $x,y
\le \frac{1}{e}$.  Because of $|h'(x)| \le 1$ on $[\frac{1}{e},1]$ and the
mean value theorem, it holds that
\begin{equation}
\frac{1}{e}\le x,y \le 1 \; \Rightarrow \; |h(x)-h(y)| \le |h'(x)||x-y| = |x-y|.
\end{equation}
Because of $(*)$ we obtain the claim for the case 
$\frac{1}{e}\le x,y \le 1$.\\
It remains the case (w.l.o.g.~$x<y$) $x<y\le\frac{1}{e}$. 
Here it holds that $|h(x)-h(y)| = h(y)-h(x)$.
We note that the function
$\log\frac{1}{t}-1$ is positive and monotonically decreasing on
$[0,\frac{1}{e}]$ $(**)$. 
We obtain the claim from the calculation
\begin{equation}
\begin{split}
|&h(x)-h(y)| = \int^y_x(\log\frac{1}{t}-1) dt
\stackrel{(**)}{\le} \int^y_x(\log\frac{1}{t-x}-1)dt\\ 
&\stackrel{s=t-x}{=} \int^{y-x}_0(\log\frac{1}{s}-1)ds
= \left[s\log\frac{1}{s}\right]^{y-x}_0 = h(y-x).
\end{split}
\end{equation}
\qed\\

Let now
$\Delta^{n-1}_K:=K\cdot\Delta^{n-1}=\{x=(x_1,...,x_n)\in\R^n\,|\,x_i\ge
0,\sum_ix_i = K\}$.  In a way that is completely analogous to that of showing
that entropy attains a maximum at uniform distributions we infer that, on
$\Delta_K^{n-1}$, the function
$h_{K,n}(x_1,...,x_n):=\sum_{i=1}^nx_i\log\frac{1}{x_i}$ (a scaled version of
entropy) attains a global maximum at $\bar{x}:= (K/n,...,K/n)$ $(***)$.

\medskip
\noindent We are now able to prove lemma~\ref{l.hnestimation}.

\medskip
\Pf
Obviously $H^t(P) = H^t(Q)$ in case of $d_{TV,t}(P,Q)=0$. In case
of $d_{TV,t}(P,Q) > 0$ 
\begin{equation}
\begin{split}
|&H^t(P) - H^t(Q)|\\
&\le \frac1t\sum_{v\in\Sigma^t}|P(v)\log\,\frac{1}{P(v)} 
                                                      - Q(v)\log\,\frac{1}{Q(v)}|\\
&\stackrel{Subl.~\ref{sl.xlnx}}{\le} \frac1t\sum_{v\in\Sigma^t}|P(v) - Q(v)|
                                     \log\,\frac{1}{|P(v) - Q(v)|}\\
&\stackrel{(***)}{\le}\frac1t\sum_{v\in\Sigma^t}\frac{d_{TV,t}(P,Q)}{|\Sigma|^t}
                                                                   \log\,\frac{|\Sigma|^t}{d_{TV,t}(P,Q)}\\
&= \frac1t d_{TV,t}(P,Q)(t\log\,|\Sigma| 
                      + \log\,\frac{1}{d_{TV,t}(P,Q)}).
\end{split}
\end{equation}
\qed\\

To get control of the limes superior resp. inferior involved in the
definition of entropy rate we will further need the following lemma.

\medskip
\begin{lemma}\label{l.limsup}
Let $(a_t)$ and $(b_t)$ two non-negative real valued sequences such that
\begin{equation}
|a_t - b_t| \le c_t \quad\text{ and }\quad \lim_{t\to\infty}c_t = c.
\end{equation}
Then it holds that
\begin{eqnarray}
\label{eq.limsup}|\limsup_{t\to\infty}a_t - \limsup_{t\to\infty}b_t| & \le & c\\
\label{eq.liminf}|\liminf_{t\to\infty}a_t - \liminf_{t\to\infty}b_t| & \le & c.
\end{eqnarray}
\end{lemma}

\Pf We only display the proof for (\ref{eq.limsup}) as that
of (\ref{eq.liminf}) can be obtained, mutatis mutandis, by analogous
considerations.\par W.l.o.g.~assume $a:= \limsup a_t \ge \limsup b_t
=: b$. Choose a subsequence $k(t)$ such that
$\lim_{t\to\infty}a_{k(t)} = a$. We obtain
\begin{equation}
\begin{split}
a - b &\le a - \limsup_{t\to\infty}b_{k(t)} = 
\limsup_{t\to\infty}a_{k(t)} - \limsup_{t\to\infty}b_{k(t)}\\
&\le \limsup_{t\to\infty}|a_{k(t)}-b_{k(t)}|\le c.
\end{split}
\end{equation}
\qed\\

We are now in position to prove theorem~\ref{t.lipschitzentropy}.
\medbreak

\Pf
As Lipschitz continuity is a local property, we can assume that
$d_{TV}(P,Q)\le\frac1e$.  Setting $a_t := H^t(P)$ and $b_t := H^t(Q)$ we
obtain by lemma~\ref{l.hnestimation} 
\begin{equation}
|a_t - b_t| \le d_{TV,t}(P,Q)(\log\,|\Sigma| 
+ \frac1t\log\,\frac{1}{d_{TV,t}(P,Q)}) =: c_t.
\end{equation}
The definition of $d_{TV,t}$ 
and lemma~\ref{l.metric} lead to
\begin{equation}
\begin{split}
\lim_{t\to\infty}c_t &= \lim_{t\to\infty}d_{TV,t}(P,Q)(\log\,|\Sigma| 
                + \frac1t\log\,\frac{1}{d_{TV,t}(P,Q)})\\ 
&= d_{TV}(P,Q)\cdot\log\,|\Sigma|.
\end{split}
\end{equation}
Plugging $(a_t),(b_t)$ and $(c_t)$ into lemma~\ref{l.limsup} then
yields the desired result.\qed\\

In order to elucidate that the structure of the proof strongly depends on the
choice of the norm we rephrase lemma~\ref{l.hnestimation} in a more general
fashion, without the ``soul'' of an entropy. Therefore let
\begin{equation}
h_n(x_1,...,x_n)=\frac{1}{\log n}\sum_{i=1}^nx_i\log \frac{1}{x_i}
\end{equation}
on $\Delta^{n-1}$
where $n\ge 2$ and $0\log\,\infty := 0$. A more prosaic version of
lemma~\ref{l.hnestimation} then reads
\begin{equation}
|h_n(x)-h_n(y)| \le ||x-y||_1 \cdot (1 + \frac{1}{\log\,n}\log\,\frac{1}{||x-y||_1}),
\end{equation}
where $||x||_1 = \sum_i|x_i|$ as usual. A straightforward consequence of the lemma
is 
\begin{multline}
\label{eq.equi}
\forall\epsilon\in\R_+\;\exists\delta\in\R_+\;\forall n\ge 2\;\forall x,y\in\Delta^{n-1}:\\
||x-y||_1 < \delta \; \Longrightarrow \; |h_n(x)-h_n(y)| < \epsilon.
\end{multline}
After being translated back to entropies, this states that entropy rate
is uniformly continuous on $\cP$. We note that the statement of
the generalized lemma need not be true relative to norms $||.||_p$
different from $||.||_1$.  More formally:

\medbreak

\begin{lemma}
\label{l.contranorm}
Let $2 \le p < \infty$ and $||x||_p= \sqrt[p]{\sum_i|x_i|^p}$ the usual
$p$-norm on $\R^n$. Then it holds that
\begin{equation*}
\exists\epsilon\in\R_+\;\forall\delta\in\R_+\;\exists N\ge 2\;\exists x,y\in\Delta^{N-1}:\quad
||x-y||_p < \delta,\;|h_N(x)-h_N(y)| > \epsilon
\end{equation*}
which is just the negation of (\ref{eq.equi}).
\end{lemma}

For the proof we use the notation ($0<m\le n$)
\begin{equation}
x^*_{m,n}:=
(\underset{m\;times}{\underbrace{\frac{1}{m},...,\frac{1}{m}}},0,...,0)\in\Delta^{n-1}.
\end{equation}

\Pf
Choose $\epsilon = 1/2$ and $\delta\in\mathbb{R}^+$ arbitrarily. Choose
an $m\in\mathbb{N}$, such that $m>\frac{1}{\delta}$. Then find an $N_0
> m$, such that $||x^*_{n,n}||_2=(\frac{1}{n})^{1/2}<\delta$ for every
$n\ge N_0$. Further
\begin{equation}
\begin{split}
||&x^*_{m,n}-x^*_{n,n}||_p \le ||x^*_{n,n}||_p 
= (\frac{1}{n^{p-1}})^{\frac{1}{p}} = n^{-\frac{p-1}{p}}\\ 
&= n^{\frac{1}{p}-1} \le n^{-\frac{1}{2}} =  ||(\frac{1}{n},...,\frac{1}{n})||_2 < \delta,
\end{split}
\end{equation}
but
\begin{equation}
|h_n(x^*_{m,n}) - h_n(x^*_{n,n})| 
= \frac{1}{\log\,n}|\log\,m - \log\,n| \underset{n\to\infty}{\longrightarrow} 1.
\end{equation}
Therefore, we find an $N\in \mathbb{N}$ and suitable $x,y\in\Delta^{N-1}$ 
which support the statement of the lemma.
\qed\\

\begin{Rem} 
Because of lemma~\ref{l.contranorm}, one could intuitively be led to the
assumption that entropy rate need not be continuous with respect to the norms
given through the spaces $L_p(\Omega,\cB,P), p\ge 2$ . However, this is not
true, see \cite{Schoenhuth07c} for respective considerations.
\end{Rem}

\section{Entropy rate of sources with finite evolution dimension}
\label{sec.finitedim}

\medskip
In the following we will give a direct proof of the existence of entropy rate
of sources with finite evolution dimension which had been introduced
in \cite{Faigle07}. See the subsequent subsection \ref{ssec.finiteexamples}
for prevalent examples of random sources of finite evolution dimension.\par
As sources with finite evolution
dimension are asymptotically mean stationary \cite{Faigle07}, the result can
be obtained as a corollary of the theorem of Shannon-McMillan-Breiman for
asymptotically mean stationary sources \cite{Gray80}. However, the following
proof is much simpler. See subsection~\ref{ssec.comparison} for a detailed
comparison of the two proofs.

\subsection{Preliminaries}
\label{subsec:finitedim}

In the following let the {\em shift operator} $T:\Omega\to\Omega$ be defined
by
\begin{equation}
T(v_0v_1v_2...) := v_1v_2...\;.
\end{equation}
Obviously, $T$ is measurable. If $(X_t)$ is a
discrete random source with associated measure $P_X$ then
\begin{equation}
(P_X\circ T^{-k})(v):=\sum_{w\in\Sigma^k}P_X(wv)
\end{equation}
gives rise to a probability measure $P\circ T^{-k}$ which is associated with the discrete
random source $((X_k)_t)$ defined through
\begin{multline*}
P_{X_k}(\{(X_k)_0=v_0,(X_k)_1=v_1,...,(X_k)_{t-1}=v_{t-1}\})\\ 
:= P_X(\{X_k=v_0,X_{k+1}=v_1,...,X_{t-1+k}=v_{t-1}\}).
\end{multline*}
A discrete random source $(X_t)$ is said to be of {\em finite evolution
dimension} if the family $(P_X\circ T^{-k})_{k\ge 0}$ spans a
finite-dimensional subspace in the linear space of finite, signed measures on
$(\Omega,\cB)$ (see appendix~\ref{app.totvarnorm} for the definition of a
finite, signed measure).

In the following we will write
\begin{equation}
PT^{-i}:=P\circ T^{-i}\quad\text{and}\quad P_n :=
\frac1n\sum_{i=0}^{n-1}PT^{-i}
\end{equation}
for probability measures $P$ associated with random sources.

\medskip
\begin{theorem}
\label{t.finitetv}
If $P$ is a discrete random source of finite evolution dimension there is a
stationary discrete random source $\bar{P}$, called the {\em stationary mean}
of $P$ such that
\begin{equation}
\lim_{n\to\infty}d_{TV}(P_n,\bar{P}) = 0.
\end{equation}
\end{theorem}

\Pf
The proof is centered on an elementary fact from linear algebra.
As it requires some of the basic theory of finite, signed measures,
we have deferred it to sec.~\ref{ssec.finitetv} in the appendix. Note
that an alternative, slightly more complicated version of the proof
has already been given in \cite{Faigle07}.
\qed\\

\subsection{Proof for the existence of entropy rate}
\label{subsec:existence}

In order to be prepared for the proof we provide a lemma whose immediate
consequence is that entropy rate coincides for all $P_n,n\ge 0$.

\medskip
\begin{lemma}
\label{l.amsinvariant}
Let $P$ be a probability measure associated with a random source.
Then it holds that
\begin{equation}
\forall n\in\mathbb{N}:\quad \lim_{t\to\infty}(H^t(P) - H^t(P_n)) = 0.
\end{equation}
\end{lemma}

\Pf 
A straightforward consequence of Lemma 2.3.4, \cite{Gray90} is that
for $\alpha\in[0,1]$ and probability measures $P,Q$:
\begin{multline*}
\alpha H^t(P) + (1-\alpha)H^t(Q)\le H^t(\alpha P+(1-\alpha)Q)\\
\le\alpha H^t(P) + (1-\alpha)H^t(Q) +  \frac{\log\,2}{t}
\end{multline*}
Now, by induction on $n$,
\begin{equation*}
\frac{1}{n}\sum_{i=0}^{n-1}H^t(PT^{-i}) \le H^t(P_n) 
\le\frac{1}{n}\sum_{i=0}^{n-1}H^t(PT^{-i}) + \frac{n}{t}\log\,2
\end{equation*}
and the assertion follows from lemma~\ref{l.entrateinvariant} 
(appendix~\ref{app.peqpcirct}) which states that the $H^t(PT^{-i})$
conincide for all $i\ge 0$.
\qed\\

We establish that both upper entropy rate $\ovH$ and lower entropy
rate $\unH$ coincide for all $P_n,n\ge 0$.

\medskip
\begin{corollary}\label{c.entrateident}
Let $P$ be the probability measure associated with a random source. Then it
holds that
\begin{equation}
\ovH(P) = \ovH(P_n)\quad\text{and}\quad
\unH(P) = \unH(P_n)
\end{equation}
for all $n\in\N$.
\end{corollary}

\Pf Use lemma~\ref{l.amsinvariant} in order to
apply lemma~\ref{l.limsup} to the sequences $(a_t:=H^t(P)),
(b_t:=H^t(P_n))$ for the first equation. For the second one
rephrase lemma~\ref{l.limsup} with $\liminf$ instead of
$\limsup$. 
\qed\\

As a consequence, we can prove the existence of entropy rate for
finite-evolution-dimensional sources.

\medskip
\begin{theorem}[Existence of entropy rate]
\label{t.finiteentropy}
Let $P$ be a probability measure associated with a random source of finite
evolution dimension. Let $\bar{P}$ be the stationary mean of $P$. Then it
holds that
\begin{equation}
\ovH(P) = \unH(P) = \lim_{t\to\infty}H^t(P).
\end{equation}
Therefore, entropy rate of $P$ exists. Moreover, it is equal to the one of
the stationary mean $\bar{P}$.
\end{theorem}

\Pf As the $P_n$ converge in TV-norm to $\bar{P}$
(theorem~\ref{t.finitetv}) we obtain due to the continuity of
$\ovH,\unH$ (theorem~\ref{t.lipschitzentropy})
\begin{equation}
\lim_{n\to\infty}\ovH(P_n) = \ovH(\bar{P})\quad\text{and}\quad 
\lim_{n\to\infty}\unH(P_n) = \unH(\bar{P}).
\end{equation}
It follows, as $\ovH(P_n)$ and $\unH(P_n)$ are constant with
respect to $n$ (corollary~\ref{c.entrateident}) and 
$\ovH(\bar{P}) = \unH(\bar{P})$ (as entropy rate exists for stationary
sources) that $\ovH(P) = \unH(P) = \ovH(\bar{P})$.
\qed\\

\begin{Rem}
Theorem \ref{t.finitetv} can be generalized to general asyptotically mean
stationary (AMS) sources (see \cite{Gray80} for the theory of AMS
sources). However, the proof needs a sophisticated
ergodic theorem, thereby loosing the elementary flavour \cite{Schoenhuth07a}.
\end{Rem}

\subsection{Examples of discrete random sources of finite evolution dimension}
\label{ssec.finiteexamples}

In the following, we present two classes of discrete random sources
that have finite evolution dimension.\par
\medskip

\paragraph{Hidden Markov Sources (HMSs)}
{\em Hidden Markov sources (HMSs)} are the discrete random sources
associated with hidden Markov models (HMMs) (also termed hidden Markov
chains in the related literature). HMSs have been largely studied, see
e.g.~\cite{Merhav02} for a comprehensive review. In the following, we
will give a brief definition of HMMs.\par An HMM $\cM=(\Sigma,S,\pi,A,E)$
is specified by a finite set of output symbols $\Sigma$, a set of hidden
states $S=\{1,...,n\}$, a transition probability matrix
$A=(A_{ij})_{i,j\in S}\in\R^{n\times n}$, an initial probability
distribution $\pi\in\R^n$ and an emission probability matrix $E =
(E_{iv})_{i\in S,v\in\Sigma}\in\R^{n\times\Sigma}$.  It gives rise to a
discrete random source $p_{\cM}$ with values in the finite set $\Sigma$,
referred to as {\em hidden Markov source} (HMS) by the idea of
changing hidden states according to the transition probabilities
$A_{ij}=P(i\to j)$, where the first state is picked according to
$\pi$, and emitting symbols from the hidden states, as specified by
the emission probabilities $E_{ia} = P(a\text{ is emitted from
}i)$. More formally, in accordance with
(\ref{eq.stringfunction}),
\begin{equation}
\label{eq.hmm}
p_{\cM}(v=v_1...v_t) = \sum_{i_1...i_t\in
  S^t}\pi(i_1)E_{i_1v_1}A_{i_1i_2}E_{i_2v_2}\cdot\hdots\cdot
A_{i_{t-1}i_t}E_{i_tv_t}.
\end{equation}
In the literature, HMSs are often introduced as being induced by {\em
  finite functions of Markov chains} where emission probability
distributions are replaced by a finite function $f:S\to \Sigma$ mapping
hidden states to output symbols. It is straightforward to see that
they give rise to complete class of HMSs as well.\par It is well known
that HMSs have {\em finite dimension} or, equivalently, have {\em
  finite degree of freedom}. See \cite{Heller65} for an early work on
the topic and \cite{Ito92} for further related work.  The relationship
of finite dimension and finite evolution dimension has been thoroughly
discussed in \cite{Faigle07}. It holds that finite evolution dimension
is a necessary condition of finite dimension, which establishes that
HMSs are of finite evolution dimension. Examples for which the 
generalization of the existence of entropy rate of sources with
finite evolution dimension apply are non-stationary HMSs. A simple
example for this might be a binary-valued source (i.e.~$\Sigma=\{0,1\}$)
induced by a ``circular'' HMM acting
on three hidden states
$S:=\{1,2,3\}$ with transition resp.~ emissionn probability matrix
\begin{equation}
\begin{bmatrix}
0 & 1 & 0\\
0 & 0 & 1\\
1 & 0 & 0
\end{bmatrix}
\quad\text{resp.}\quad
\begin{bmatrix}
0 & 1\\
0.5 & 0.5\\
1 & 0
\end{bmatrix}.
\end{equation}
Clearly, this source is not stationary such that the simple existence
proof for stationary sources does not apply. However, as an HMS, this
source is of finite evolution dimension such that theorem
\ref{t.finiteentropy} ensures the existence of its entropy rate.\par
See the subsequent sec.~\ref{ssec.comparison} for a comparison of
available proofs of the existence of entropy rate.
\medskip

\begin{Rem}
Related work on analytic properties of entropy rate of HMSs is
concerned with topologies referring to the parameterizations of the
HMMs giving rise to the HMSs, that is, with the natural topologies of
real-valued vector spaces (e.g.~\cite{Marcus06,Marcus07,Marcus08}).
For example, in the special case of binary valued i.i.d.~processes,
emitting values from $\{0,1\}$, entropy rate is computed as
\begin{equation}
p\log(\frac1p) 
\end{equation}
where the only parameter $p$ is the probability that the binary valued
i.i.d.~process emits a $1$. Clearly, $p\log(1/p)$ is not Lipschitz
continuous in intervals around zero [$(p\log(1/p))' = \log
  1/p-1$]. However, this does not contradict theorem
\ref{t.lipschitzentropy} as convergence w.r.t.~the parameterization
does not imply convergence w.r.t.~the norm of total variation, which
we will briefly outline in the following.\par As follows from
elementary measure theoretical considerations, the topology induced by
the norm of total variation is equivalent to that of the general
version of the {\em metric of total variation}
\begin{equation}
\label{eq.totvarmetric}
D_{TV}(P,Q) := \sup_{B\in\cB}|P(B)-Q(B)|
\end{equation}
where $P,Q\in\cP$ are two probability measures acting on the
measurable space $(\Omega,\cB)$. In the case of the measurable
sequence spaces under consideration here, the equivalence of the
topologies of the metric and the norm of total variation can be seen
by lemma \ref{l.metric} as it follows from straightforward elementary
computations that the topologies of $d_{TV}$ of lemma \ref{l.metric}
and the metric of total variation $D_{TV}$ of (\ref{eq.totvarmetric})
are equivalent.  As a consequence, convergence in the sense of the
norm of total variation is equivalent to uniform convergence on all
measurable sets, that is,
\begin{equation}
\label{eq.normequiv}
\lim_{n\to\infty}||P-P_n||_{TV} = 0\quad\Leftrightarrow\quad
\lim_{n\to\infty}\sup_{B\in\cB}|P(B)-P_n(B)| = 0
\end{equation}
where $P,P_n\in\cP$.\par However, as outlined in \cite{Marcus06},
sec.~VIII, convergence of probability measures induced by hidden
Markov models whose parameterizations converge may not even be {\em
  strong} (see \cite{Jacka97} for definitions and characterizations of
several forms of convergence of probability measures) meaning that
there might exist a set $B^*\in\cB$ for which
\begin{equation}
\limsup_{n\to\infty}|P(B^*)-P_n(B^*)| > 0
\end{equation}
where the $P_n,n=0,1,...$ are hidden Markov models whose
parameterizations converge to the parameterization of $P$. According
to (\ref{eq.normequiv}), this means that convergence in terms of the
parameterization does not necessarily imply convergence w.r.t.~the
norm of total variation.
\end{Rem}

\medskip

\paragraph{Quantum Random Walks (QRWs)}
{\em Quantum random walks (QRWs)} were introduced to quantum
information theory in 2001 as an analogon to classical Markov sources
\cite{Aharonov01}.  For example, they allow to emulate Markov Chain
Monte Carlo approaches on quantum computers. However, their properties
are much less understood.  A QRW $\cQ=(G,U,\psi_0)$, in a very general
form (see \cite{Aharonov01} for the full range of definitions), is
specified by a directed, $K$-regular graph $G=(V,E)$, a unitary ({\em
  evolution}) operator $U:\C^N\to\C^N$ and a {\em wave function}
$\psi_0\in\C^N$ (i.e.~$||\psi_0||=1$ for $||.||$ the Euclidean norm)
where $N:=K\cdot |V|=|E|$. Dimensions are labeled by edges which in
turn are labeled by $(u,x)$ where $u\in V$ and $x\in X,|X|=K$ and
$\C^N$ is considered to be spanned by the orthonormal basis
$(e_{(u,x)})_{(u,x)\in V\times X=E}$.  A QRW induces a classical
random source $p_{\cQ}$ with values in $\Sigma:=V$ (i.e.~the set of
nodes) by the following iterative procedure. In the first step, the
evolution operator is applied to the initial wave function $\psi_0$,
and the resulting wave function $U\psi_0$, with probability
$\sum_{x\in X}|(U\psi_0)_{(u_1,x)}|^2$, is collapsed (i.e.~projected
and renormalized, which models a quantum mechanical measurement) to
the subspace of $\C^N$, spanned by the vectors $e_{(u_1,x)},x\in X$
that is associated with (the edges leaving from) node $u_1$, thereby
generating the first symbol $u_1$. This procedure results in a new
wave function describing the state $\psi_{u_1}$ the QRW is in after
having generated the first symbol $u_1$. In order to generate a second
symbol $U$ is applied to $\psi_{u_1}$, and $U\psi_{u_1}$ is, with
probability $\sum_{x\in X}|(U\psi_{u_1})_{(u_2,x)}|^2$, collapsed to
state $\psi_{u_1u_2}$, thereby generating the second symbol
$u_2$. Iterative application of this basic procedure of evolving
followed by collapsing yields a sequence of symbols. See
\cite{Aharonov01} for further details.\par A concise formal
description in terms of formula analogous to (\ref{eq.hmm}) of the
discrete random source $p_{\cQ}$ along with a proof of QRWs being of
finite dimension has been presented in \cite{Schoenhuth08}. Finite
evolution dimension follows from finite dimension, which, as outlined
above, has been thoroughly discussed in \cite{Faigle07}.

\subsection{Comparison of existence proofs of entropy rate}
\label{ssec.comparison}

The result of theorem~\ref{t.finiteentropy} for the special case of
HMSs can be obtained as a combination of the Shannon-McMillan-Breiman
(SMB) theorem for {\em asymptotically mean stationary (AMS)} sources
\cite{Gray80} and the fact that HMSs are AMS \cite{Kieffer81} (see
also \cite{Merhav02} for a comprehensive review of theoretical results
on HMSs). Therefore, the existence of entropy rate for arbitrary,
stationary and non-stationary, HMSs has theoretically been known since
1981. For QWRs the result has been known since 2006, implied by
combining the results of \cite{Gray80} and \cite{Faigle06} in the same
fashion as for HMSs.  However, even for HMSs, the result seems to be
rather unnoticed which might be due to both the complex nature of its
proof and that the necessary combination of results has not been
explicitly mentioned.  The SMB theorem in this most generalized
version is centered around a proof for the class of ergodic,
stationary random sources
\cite{Shannon48,McMillan53,Breiman57,Breiman60} which requires
involved ergodic theorems. The extension to general stationary sources
\cite{Billingsley3,Jacobs59,Jacobs62}, in an exemplary (and elegant)
version, needs the sophisticated concept of the ergodic decomposition
of stationary random sources \cite{Gray74}. The final step
\cite{Gray80} requires again a collection of non-trivial theorems as a
prerequisite.\\ The proof given here is substantially simpler from two
main aspects. First, it is centered around the standard elementary
proof of the existence of entropy rate of stationary sources. Note
that, this way, we do not even need to introduce ergodicity. Second,
the extension to non-stationary classes of random sources is done by
results of exclusively elementary nature (theorems~\ref{t.finitetv},
\ref{t.lipschitzentropy}).

\section{Conclusion}
\label{sec.conclusion}
We show that entropy rate is Lipschitzian relative to the topology of total
variation in an elementary fashion. Besides from providing a comparatively
simple existence proof for HMSs and QRWs, this helps getting a more
general grip of entropy rate. Moreover, it brings up some interesting open
questions:
\begin{itemize}
\item A first open question which immediately arises is whether our arguments
can be strengthened to stricter analytic properties. A first clue is that the
definition of entropy rate as well as theorem~\ref{t.lipschitzentropy} can be
consistently extended to the whole real vector space of finite, signed
measures. Rademacher's theorem \cite{Federer} states that Lipschitzian
functionals on finite-dimensional real vector spaces are differentiable almost
everywhere w.r.t.~the Lebesgue measure on the Borel-sets. This points at that
entropy rate is close to being differentiable and, so far, we have not
succeeded in constructing a random source at which entropy rate is not
differentiable.
\item The intuition behind our proof is that entropy rate cannot differ too
much if sets of typical sequences of two sources overlap to a sufficiently
high degree. However, it seems to be obvious that entropy rate is continuous
when considering it relative to sizes of sets of typical sequences which is a
more general assumption. A corresponding result would certainly be applicable
to coarser topologies as, say, the weak topology. So far, it has only been
known that entropy rate, as a functional on the set of stationary sources
only, is upper semicontinuous \cite{Gray90} relative to the weak topology. We
believe that theorems of the quality of theorem~\ref{t.lipschitzentropy},
based on the comparison of sizes of sets of typical sequences, will greatly
improve such results.
\end{itemize}

\begin{appendix}

\section{The norm of total variation}
\label{app.totvarnorm}

In the following, let $A\,\dot{\cup}\,B$ be the disjoint union
of two sets $A$ and $B$ and $\complement A$ be the 
complement of a set $A$.

\subsection{Finite signed measures}

A finite, signed measure on $(\Omega,\cB)$
is a $\sigma$-additive but not necesarily positive, finite
set function on $\cB$. The
most important relevant properties of finite signed measures are
summarized in the following theorem (see \cite{Halmos}, ch.~VI for
proofs).

\medbreak
\begin{theorem}\label{t.jordan}\hfill\\[-1ex]
\begin{enumerate}
\item By eventwise addition and scalar multiplication, the set of finite
signed measures can be considered as a real-valued vector space.
\item The \emph{Jordan decomposition} theorem states that for every $P\in\cP$
there are finite measures $P_+,P_-$ such that
\begin{equation}
P = P_+ - P_-
\end{equation}
and for all other decompositions $P = P_1 - P_2$ with measures
$P_1,P_2$ it holds that $P_1 = P_+ + \delta, P_2 = P_- + \delta$
for another measure $\delta$. In this sense, $P_+$ and $P_-$ are
unique and called {\em positive} resp. {\em negative variation}.
The measure $|P|:=P_+ + P_-$ is called {\em total variation}.

\item In parallel to the Jordan decomposition we have the \emph{Hahn
decomposition} of $\Omega$ into two disjoint events
$\Omega_+,\Omega_-$
\begin{equation}
\Omega = \Omega_+ \;\dot{\cup}\; \Omega_-
\end{equation}
such that $P_-(\Omega_+) = 0$ and $P_+(\Omega_-) = 0$.
$\Omega_+,\Omega_-$ are uniquely determined up
to $|P|$-null-sets.

\item The \emph{norm of total variation} $||.||_{TV}$ on
$\cP$ is given by
\begin{equation}
||P||_{TV} :=|P|(\Omega) = P_+(\Omega) + P_-(\Omega)\\ = P_+(\Omega_+) + P_-(\Omega_-).
\end{equation}
Obviously $||\,|P|\,||_{TV} = ||P||_{TV}$.
\end{enumerate}
\end{theorem}

\subsection{Proof of lemma \ref{l.metric}}\label{ssec.metricproof}
For the proof, we will identify
cylinder sets $B\in\cB$ with sets of words $A_B\in\Sigma^t$ as usual
($B$ is the set of sequences which are the continuations of the words
in $A_B$).  In our notation, we correspondingly obtain
\begin{equation}\label{eq.signedcylinder}
P(B) = \sum_{v\in A_B}P_+(v)-P_-(v)
\end{equation}
for a signed measure $P$ with Jordan decomposition $P=P_+-P_-$. We
will further make use of the approximation theorem (see Halmos
\cite{Halmos}, p.~56, Th.~D) which tells that, given a measure $P$, an
event $B\in\cB$ and $\epsilon\in\R_+$, we find a cylinder set $F$ such
that
\begin{equation}
P(B\;\triangle\; F) < \epsilon,
\end{equation}
where $B\triangle F = (B\setminus F) \cup (F\setminus B)$ is the
symmetric set difference.  A straightforward consequence of this is
that $|P(B)-P(F)|<\epsilon$.\\

\Pf
It suffices to show
\begin{equation}
||P||_{TV} = \sup_{t\in\N}\sum_{v\in\Sigma^t}|P(v)| 
           = \lim_{t\to\infty}\sum_{v\in\Sigma^t}|P(v)|.
\end{equation}
for an arbitrary finite, signed measure $P$.  

The second equation of (\ref{eq.totvar}) now follows immediately from 
\begin{equation}
\sum_{v\in\Sigma^t}|P(v)| 
= \sum_{v\in\Sigma^t}\underset{=|P(v)|}{\underbrace{|\sum_{a\in\Sigma}P(va)|}} 
\le \sum_{v\in\Sigma^t}\sum_{a\in\Sigma}|P(va)| 
= \sum_{v\in\Sigma^{t+1}}|P(v)| 
\end{equation}
which shows that $(\sum_{v\in\Sigma^t}|P(v)|)_{t\in\N}$ is
a monotonically increasing sequence. It remains to show that it
converges to $||P||_{TV}$. This translates to
demonstrate that, given
$\epsilon\in\R_+$, there is $T_0\in\N$ with
\begin{equation}
\sum_{v\in\Sigma^{T_0}}|P(v)| > ||P||_{TV} - \epsilon.
\end{equation}

Therefore let $P_+,P_-$ be the Jordan decomposition of $P$ and,
correspondingly, $\Omega = \Omega_+ \dot{\cup}\, \Omega_-$ be the Hahn
decomposition. By an application of the approximation theorem (see
above) we find $T_0\in\N$ and a cylinder
set corresponding to $A\subset\Sigma^{T_0}$ with
\begin{equation}\label{eq.omegaplus}
|P|(\Omega_+\;\triangle\; A) < \frac{\epsilon}{4}
\end{equation} 
a straightforward ($|P| = P_+ + P_-$) 
consequence of which is that both
\begin{equation}\label{eq.cons1}
P_+(\Omega_+\;\triangle\; A) < \frac{\epsilon}{4}
\quad\text{ and }\quad
P_-(\Omega_+\;\triangle\; A) < \frac{\epsilon}{4}
\end{equation}
Now note that the obvious
$\complement A\;\triangle \;\complement B = A\;\triangle\; B$
in combination with  $\Omega_- = \complement \Omega_+$ and
(\ref{eq.cons1}) yields
\begin{equation}\label{eq.omminus}
P_-(\Omega_- \;\triangle\;\complement A) = P(\Omega_+\;\triangle\; A) < \frac{\epsilon}{4}.
\end{equation}
(\ref{eq.cons1}) and (\ref{eq.omminus}) then yield the inequalities 
\begin{equation}\label{eq.ineq1}
P_+(\complement A) \stackrel{P_+(\Omega_-) = 0}{=} P_+(\Omega_+\setminus A)
\le P_+(\Omega_+\;\triangle\; A) < \frac{\epsilon}{4}
\end{equation}
and
\begin{equation}\label{eq.ineq2}
P_-(A) \stackrel{P_-(\Omega_+) = 0}{=} P_-(\Omega_-\setminus\complement A)
\le P_-(\Omega_-\;\triangle\;\complement A) < \frac{\epsilon}{4}.
\end{equation}
Moreover, it is straightforward from (\ref{eq.cons1}) and
(\ref{eq.omminus}) that
\begin{equation}\label{eq.ineq3}
P_+(A) > P_+(\Omega_+) - \frac{\epsilon}{4}\quad\text{and}\quad
P_-(\complement A) > P_-(\Omega_-) - \frac{\epsilon}{4}.
\end{equation}
We finally compute
\begin{equation}
\begin{split}
\sum_{v\in\Sigma^{T_0}}&|P(v)| 
= \sum_{v\in A}|P(v)| 
+ \sum_{v\in\complement A}|P(v)| \\
&\ge |P(A)| + |P(\complement A)|\\ 
&\ge P_+(A) - P_-(A) + P_-(\complement A) - P_+(\complement A)\\ 
&\overset{(\ref{eq.ineq1}),(\ref{eq.ineq2})}{\underset{(\ref{eq.ineq3})}{>}} 
(P_+(\Omega_+) - \frac{\epsilon}{4}) - \frac{\epsilon}{4} 
+ (P_-(\Omega_-) - \frac{\epsilon}{4}) - \frac{\epsilon}{4}\\
&= P_+(\Omega_+) + P_-(\Omega_-) - \epsilon = ||P||_{TV} - \epsilon.
\end{split}
\end{equation}
\qed\\

\subsection{Proof of theorem \ref{t.finitetv}}
\label{ssec.finitetv}
We start with the following lemma.
\medskip

\begin{lemma}
\label{l.ttv}
Let $P$ be a finite signed measure on $(\O,\cB)$ and $T:\O\to\O$
a measurable function. Then $P\circ T^{-1}$ is a finite signed measure
for which
\begin{equation}
|P\circ T^{-1}|(B) \le |P|(T^{-1}B)
\end{equation}
for all $B\in\cB$. In particular, 
\begin{equation}
\label{eq.mustable}
||P\circ T^{-1}||_{TV} \le
||P||_{TV}.
\end{equation}
\end{lemma}
\medskip

\Pf
Note that $P\circ T^{-1} = P_+\circ T^{-1} - P_-\circ T^{-1}$ is a
decomposition into a difference of measures. Because of the uniqueness
property of the Jordan decomposition (see th.~\ref{t.jordan}), there
is a measure $\delta$ such that $P_+\circ T^{-1} = (P\circ T^{-1})_+
+\delta$ and $P_-\circ T^{-1} = (P\circ T^{-1})_- + \delta$. Therefore
$|P\circ T^{-1}|(B) = (P\circ T^{-1})_+(B) + (P\circ T^{-1})_-(B)\le
P_+(T^{-1}B) + P_-(T^{-1}B) = |P|(T^{-1}B)$. $B=\O$ yields the last
assertion, as $T^{-1}\O = \O$.
\qed\\

{\it Proof of Th.~\ref{t.finitetv}.$\;\;$}
We recall that, in th.~\ref{t.finitetv}, $T$ was supposed to be the
shift operator, which is measurable. We observe that
\begin{equation}
\mu P := (P\circ T^{-1})
\end{equation}
establishes a linear operator on the vector space of finite, signed
measures. Due to lemma \ref{l.ttv}, (\ref{eq.mustable}), it holds that
$||\mu P||_{TV}\le ||P||_{TV}$ for all finite signed measures $P$, which
establishes
\begin{equation}
||\mu || \le 1
\end{equation}
where $||.||$ is the operator norm associated with the norm of total
variation.\par Now consider the subspace $\cP_P$ of the finite signed measures
spanned by all $P\circ T^{-i},i\in\N$ for a given finite signed measure.
Note that an equivalent description of finite evolution dimension is just
\begin{equation}
\dim\cP_P < \infty.
\end{equation}
Note further that
\begin{equation}
\mu(\cP_P)\subset\cP_P.
\end{equation}
The elementary, linear algebraic lemma 3.2 in \cite{Faigle07}
states that, given an endomorphism $F:V\to V$ on a finite-dimensional
real- or complex-valued vector space $V$ with $||F||\le 1$, for all
$x\in V$ there is an $F$-invariant $\bar{x}\in V$ such that
\begin{equation}
\lim_{n\to\infty}||\frac1n\sum_{k=0}^{n-1}F^kx - \bar{x}|| = 0.
\end{equation}
As all norms are equivalent on $V$, this applies for arbitrary choices of
norms $||.||$. Replacing $V$ by $\cP_P$, $||.||$ by
$||.||_{TV}$, $F$ by $\mu$ and $x$ by $P$ concludes the proof of
theorem \ref{t.finitetv}.
\qed\\

\section{Proof of lemma~\ref{l.entrateinvariant}}
\label{app.peqpcirct}

\begin{lemma}\label{l.entrateinvariant}
Let $P$ be a discrete random source. Then it holds that
\begin{equation}
\lim_{t\to\infty}(H^t(P) - H^t(P\circ T^{-k})) = 0.
\end{equation}
\end{lemma}

\Pf
Using the notation 
\begin{equation}
I^k_t(P) 
:= \frac{1}{t}\sum_{v\in\Sigma^k}\sum_{w\in\Sigma^t}P(vw)
\log\,\frac{PT^{-k}(w)}{P(vw)}
\end{equation}
and 
\begin{equation}
J^k_t(P) := 
\frac{1}{t}\sum_{v\in\Sigma^k}\sum_{w\in\Sigma^t}P(vw)
\log\,\frac{P(v)}{P(vw)}
\end{equation}
one obtains
\begin{equation}
H^t(P) + J^k_t(P) \stackrel{(*)}{=} H^{k+t}(P) 
= I^k_n(P) + H^t(P\circ T^{-k})
\end{equation}
where $(*)$ follows from a well known and elementary theorem (e.g. \cite{Han},
p.22, theorem 2.1) and the second equation is obvious.  Because of
\begin{equation}
0 \le J^k_t(P) \le \frac{k}{t}H^k(P\circ T^{-t}) 
S\le \frac{1}{t}\log\,|\Sigma^k| \underset{t\to\infty}{\longrightarrow} 0
\end{equation}
and
\begin{equation}
0 \le I^k_t(X) \le \frac{1}{t}H^k(X) 
\le \frac{1}{t}\log\, |\Sigma^k| \underset{t\to\infty}{\longrightarrow} 0,
\end{equation}
the assertion follows from an application of the sandwich theorem.
\qed\\

\end{appendix}

\section*{Acknowledgment}
I would like to thank Ulrich Faigle who considerably contributed to
the work presented here. I also would like to thank the unknown
reviewers for helpful comments and suggestions.


\begin{thebibliography}{1}

\bibitem{Aharonov01}
D. Aharonov, A. Ambainis, J. Kempe, U. Vazirani, "Quantum walks on
graphs", in \emph{Proc. of 33rd ACM STOC, New York}, 2001, pp.~50-59.

\bibitem{Billingsley3}
P.~Billingsley,
{\em Ergodic Theory and Information},
Wiley, 1965.

\bibitem{Breiman57}
L.~Breiman, ``The individual ergodic theorem of information theory'',
in {\em Annals of Mathematical Statistics}, 1957, vol.~28, pp.~809-811.

\bibitem{Breiman60}
L.~Breiman,
\newblock A correction to 'the individual ergodic theorem of information
  theory'.
\newblock {\em Annals of Mathematical Statistics}, 31:809--810, 1960.

\bibitem{Faigle06}
U. Faigle, A. Sch\"onhuth, "Quantum predictor models",
\emph{Electonic Notes in Discrete Mathematics}, 2006, vol.~25, pp.~149-155.

\bibitem{Faigle07}
U.~Faigle and A.~Schoenhuth,
``Asymptotic mean stationarity of sources with finite evolution dimension'',
{\em IEEE Trans.~Inf.~Theory}, 2007, vol.~53(7), pp.~2342-2348.

\bibitem{Federer}
H.~Federer, {\em Geometric Measure Theory}.
Springer, 1969.

\bibitem{Gray74} 
R.~Gray and L.~Davisson, ``The ergodic decomposition of
stationary discrete random processes'', {\em IEEE Transactions on Information
Theory}, 1974, vol.~20(5), pp.~625-636.

\bibitem{Gray80}
R.M. Gray and J.C. Kieffer, ``Asymptotically mean stationary measures''
{\em Annals of Probability}, 1980, vol.~8, pp.~962--973.

\bibitem{Gray90}
Robert~M. Gray, {\em Entropy and Information Theory}.
Springer Verlag, 1990.

\bibitem{Halmos}
P.R. Halmos, {\em Measure Theory}.
Van Nostrand, 1964.

\bibitem{Han}
T.S.~Han and K.~Kobayashi,
{\em Mathematics of Information and Coding}.
American Mathematical Society, 2002.

\bibitem{Heller65}
A.~Heller
``On stochastic processes derived from Markov chains'', 
{\em Annals of Mathematical Statistics}, vol.~36(4), pp.~1286-1291, 1965

\bibitem{Ito92}
H.~Ito, S.-I.~Amari and K.~Kobayashi:
``Identifiability of hidden Markov information sources and their minimum
degrees of freedom'', {\em IEEE Trans.~Inf.~Theory}, vol.~38(2), pp.~324--333, 1992.

\bibitem{Jacka97}
S.D. Jacka and G.O. Roberts, 
``On strong forms of weak convergence''
{\em Stochastic Processes and Applications}, 1997, vol.~67, pp.~41-53.

\bibitem{Jacobs59}
K.~Jacobs, ``Die Übertragung diskreter Informationen durch periodische und
  fastperiodische Kanäle'', 1959, {\em Mathematische Annalen}, vol.~137, pp.~125-135.

\bibitem{Jacobs62}
K.~Jacobs, ``Über die Struktur der mittleren Entropie'', 
{\em Mathematisches Zentralblatt}, 1962, vol.~78, pp.~33-43.

\bibitem{Jacquet04}
P.~Jacquet, G.~Seroussi, and W.~Szpankowski, ``On the entropy of a hidden markov process'',
in {\em Proc. Data Compression Conf.}, Snowbird, UT, March 2004, pp.~362-371.

\bibitem{Kieffer81}
J.C. Kieffer and M.~Rahe, ``Markov channels are asymptotically mean stationary'',
{\em SIAM J. Math. Anal.}, 1981, vol.~12(3), pp.~293-305.

\bibitem{Marcus06}
G.~Han and B.~Marcus, ``Analyticity of entropy rate of hidden Markov chains'',
{\em IEEE Trans.~Inf.~Theory}, 2006, vol.~52(12), pp.~5251-5266.

\bibitem{Marcus07}
G.~Han and B.~Marcus, ``Derivatives of entropy rate in special families of hidden Markov chains'',
{\em IEEE Trans.~Inf.~Theory}, 2007, vol.~53(7), pp.~2642-2652.

\bibitem{Marcus08}
G.~Han and B.~Marcus, ``Asymptotics of entropy rate of hidden Markov chains at weak black holes'',
{\em Proc.~IEEE Int.~Symp.~Inf.~Th.}, 2008, pp.~2629-2633.

\bibitem{McMillan53}
B.~McMillan, ``The basic theorems of information theory'',
{\em Annals of Mathematical Statistics}, 1953, vol.~24, pp.~196-219.

\bibitem{Merhav02}
Y.~Ephraim, N.~Merhav,
"Hidden Markov processes", \emph{IEEE Trans. on Information Theory},
vol.~48(6), pp.~1518-1569.

\bibitem{Ordentlich06}
E.~Ordentlich and T.~Weissman, ``On the optimality of symbol by symbol filtering and denoising'',
{\em IEEE Trans.~Inf.~Theory}, 2006, 52(1), pp.~19-40.

\bibitem{Schoenhuth07a}
A. Sch\"onhuth, 
``The ergodic decomposition of asymptotically mean stationary random sources'',
submitted manuscript, http://arxiv.org/abs/0804.2487.

\bibitem{Schoenhuth07c}
A. Sch\"onhuth, ``On analytic properties of entropy rate'',
2006, technical report, ZAIK, University of Cologne.

\bibitem{Schoenhuth08}
A. Sch\"onhuth, 
``A simple and efficient solution of the identifiability problem for hidden Markov sources
and quantum random walks'',
ISITA 2008, to appear, http://arxiv.org/abs/0808.2833.

\bibitem{Shannon48}
C.~Shannon, ``A mathematical theory of communication'',
{\em Bell System Technical Journal}, 1948.

\bibitem{Zuk06}
O.~Zuk, E.~Domany, I.~Kanter, and M.~Aizenman, 
``From finite system entropy to entropy rate for a hidden markov process'', 
{\em IEEE Signal Processing Letters}, 2006, vol.~13(9), pp.~517-520.

\bibitem{Zuk05}
O.~Zuk, I.~Kanter, and E.~Domany.
\newblock The entropy of a binary hidden markov process.
\newblock {\em Journal of Statistical Physics}, 2005, vol.~121(3-4), pp.~343-360.

\end{thebibliography}
\end{document}